\documentclass[preprintnumbers,amsmath,amssymb,aps,prl,longbibliography,floatfix,lengthcheck,superscriptaddress]{revtex4-2}
\usepackage{units}
\usepackage{graphicx}
\usepackage{braket}
\usepackage{txfonts}
\usepackage[colorlinks=true,citecolor=blue]{hyperref}
\usepackage{xcolor}
\usepackage{orcidlink}
\usepackage{pdfpages}
\usepackage{etoolbox}

\makeatletter
\patchcmd{\@outputpage@head}{\@ifx{\LS@rot\@undefined}{}{\LS@rot}}{}{}{}
\makeatother

\newcommand{\upmu}{\muup}

\begin{document}
	
	\title{Quadrupole coupling of circular Rydberg qubits to inner shell excitations}
	
	\author{M. Wirth\orcidlink{0009-0007-6940-7916}}
	\author{C. H\"{o}lzl\orcidlink{0000-0002-2176-1031}}
	\author{A. G\"{o}tzelmann\orcidlink{0000-0001-5527-5878}}
	\author{E. Pultinevicius\orcidlink{0009-0005-7404-9178}}
	\author{F. Meinert\orcidlink{0000-0002-9106-3001}}
	\affiliation{5. Physikalisches Institut and Center for Integrated Quantum Science and Technology, Universit\"{a}t Stuttgart, Pfaffenwaldring 57, 70569 Stuttgart, Germany}
	\date{\today}
	
	\begin{abstract}
		Divalent atoms provide excellent means for advancing control in Rydberg atom-based quantum simulation and computing, due to the second optically active valence electron available. Particularly promising in this context are circular Rydberg atoms, for which long-lived ionic core excitations can be exploited without suffering from detrimental autoionization. Here, we report the implementation of electric quadrupole coupling between the metastable 4\,D$_{3/2}$ level and a very high-$n$ ($n=79$) circular Rydberg qubit, realized in doubly excited $^{88}$Sr atoms prepared from an optical tweezer array. We measure the kHz-scale differential level shift on the circular Rydberg qubit via beat-node Ramsey interferometry comprising spin echo. Observing this coupling requires coherent interrogation of the Rydberg states for more than one hundred microseconds, which is assisted by tweezer trapping and circular state lifetime enhancement in a black-body radiation suppressing capacitor. Further, we find no noticeable loss of qubit coherence under continuous photon scattering on the ion core, paving the way for laser cooling and imaging of Rydberg atoms. Our results demonstrate access to weak electron-electron interactions in Rydberg atoms and expand the quantum simulation toolbox for optical control of highly excited circular state qubits via ionic core manipulation.
	\end{abstract}
	
	\maketitle
	
	Circular Rydberg atoms have a long history in experiments on the fundamental quantum nature of light-matter interactions at the level of individual atoms and photons \cite{Haroche2013}. Among groundbreaking studies are non-demolition measurements of photons \cite{Gleyzes2007,Brune2008}, stabilization of photon number states via quantum feedback \cite{Sayrin2011}, and the generation of cat states of light \cite{Deleglise2008,Zhou2012}. Key to these experiments was the long lifetime of circular Rydberg states (CRS) \cite{Hulet1983}, which is due to their maximum angular momentum ($\vert m \vert = n-1$, where $m$ and $n$ denote the orbital magnetic and principal quantum number). Atomic qubits are then defined by circular states of nearby manifolds, strongly coupled to microwave photons trapped in superconducting cavities.
	
	More recently, the long lifetime of CRS is attracting increasing interest for boosting the coherence times in neutral atom quantum computers and simulators based on arrays of optical tweezers \cite{Ma2023, Nguyen2018,Meinert2020}. While first tweezer arrays of circular states have been reported for rubidium atoms \cite{Ravon2023}, the coherent control of long-lived circular state arrays in the divalent alkaline earth metal strontium followed soon after \cite{Hoelzl2024}. Divalent atoms are particularly promising in the context of quantum simulation \cite{Cooper2018,Norcia2018}, because of the new control capabilities that come along with the optically active ionic core of the Rydberg atom.

	For example, polarizing the orbital of the second valence electron by a far off-resonant light field enables to trap the Rydberg atom in a standard Gaussian tweezer beam \cite{Hoelzl2024}, a method that is also applicable to low-angular momentum states \cite{Wilson2022}. However, unlike their low-angular momentum counterparts \cite{Cooke1978,Millen2010,Madjarov2020}, CRS even remain stable when resonantly exciting their ionic core as a result of the convenient absence of autoionization for high angular momentum orbitals \cite{Teixeira2020,Lehec2021}. The long lifetime of the core-excited Rydberg atom then allows to probe the coherent interaction between the two valence electrons, and to exploit it for quantum state control. This opens exciting perspectives for local optical manipulation of the circular Rydberg microwave qubit, fluorescence detection \cite{McQuillen2013} and even laser cooling \cite{Bouillon2024}. More specifically, a recent experiment on an atomic strontium beam revealed a strongly $n$-dependent interaction shift of hundreds of kilohertz on the CRS for $n=51$, when the ionic core is shelved into the metastable 4\,D$_{3/2}$ level \cite{Muni2022}. The shift arises from the interaction of the far-away Rydberg electron with the quadrupole moment of the ionic core orbital \cite{Safronova2008}.
	
	Here, we leverage this electric quadrupole coupling for coherent control of very high-$n$ ($n = 79$) circular Rydberg qubits prepared from an array of individual tweezer-trapped strontium atoms. The high principal quantum number allows for boosting the ratio of coherent interaction cycles over the state lifetime in a circular state quantum simulator, which is particularly interesting for room temperature setups \cite{Meinert2020,Hoelzl2024}. As the quadrupole shift rapidly decreases with increasing $n$ ($\propto n^{-6}$), we have to fully exploit the long interrogation time provided by our trapped atom array of hundreds of microseconds combined with spin-echo interferometry. Furthermore, we make use of CRS lifetime enhancement to the millisecond range in our room temperature setup due to a pair of capacitor plates that suppresses black-body radiation-induced decay \cite{Hoelzl2024,Wu2023,Hulet1985}. Finally, no noticeable loss of coherence under continuous photon scattering on the inner core is observed, which fosters coherence preserving optical detection or cooling of trapped Rydberg atoms.
	
	\begin{figure}[!ht]
		\centering
		\includegraphics[width=\columnwidth]{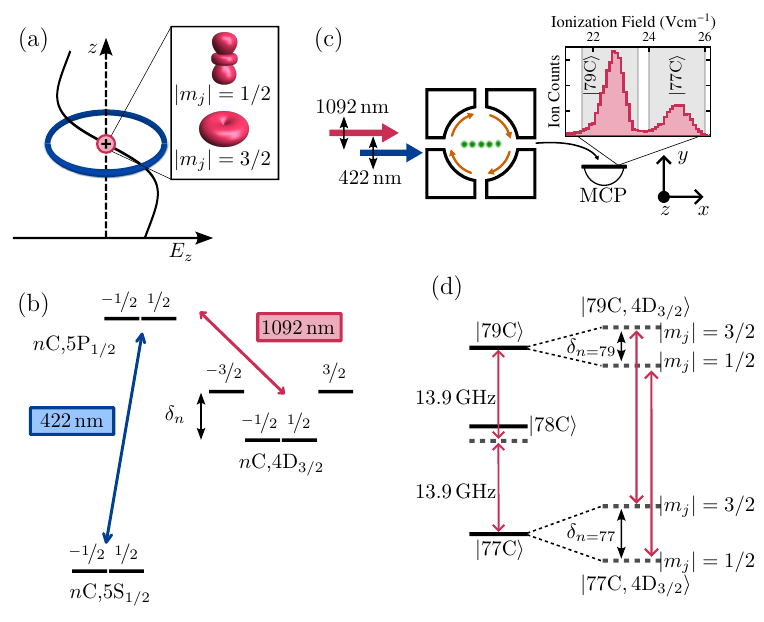}
		\caption{(a) Origin of the quadrupole interaction shift on the CRS. The quadrupole moment of the core 4\,D$_{3/2}$ electron wavefunction couples to the gradient of the electric field $E_z$ (black), that the Rydberg electron orbit (blue) generates at the core position along the quantization axis ($z$-axis). The inset shows the spin-weighted orbitals of the 4\,D$_{3/2},m_j$ substates, with oblate (prolate) shape and thus a positive (negative) quadrupole moment. (b) Level diagram of the electronic core states and their respective magnetic substates $m_j$, with the Rydberg electron in the circular orbit $|n\text{C}\rangle$. The metastable 4\,D$_{3/2}$ sublevels are split by $\delta_n$ due to their quadrupole interaction with the Rydberg electron. (c) Schematic of the experiment. Single $^{88}$Sr atoms are trapped in optical tweezers (green) inside an electrode structure consisting of four circularly shaped electrodes for applying $\sigma^+$-polarized RF fields and two black-body radiation suppression capacitor plates in the $z$-direction (not shown). Co-propagating laser beams at $\unit[422]{nm}$ and $\unit[1092]{nm}$ (blue and red arrow, with polarization along $y$ as indicated) control the ion core electron. State-selective field ionization and ion detection on an MCP enables readout of the circular Rydberg qubit. The inset shows an exemplary ion histogram with the $|79\text{C}\rangle$ and $|77\text{C}\rangle$ detection windows (gray). (d) Level scheme of the circular Rydberg qubit, driven by a two-photon microwave transition at $\unit[13.9]{GHz}$ (red arrows), including the quadrupole interaction shift $\delta_n$.}
		\label{Fig1}
	\end{figure}
	
	To start with, we consider a strontium circular Rydberg atom with the ionic core in the metastable excited 4\,D$_{3/2}$ fine structure state. The inner valence electron has magnetic substates $|m_j| = 1/2$ and $3/2$, with spin-weighted orbitals as depicted in Fig.~\ref{Fig1}(a). These orbitals possess a static electric quadrupole moment, which interacts with the Coulomb field of the distant Rydberg electron. To be more specific, the circular Rydberg electron produces an electric field gradient $\partial_z E_z$ along the $z$-direction at the position of the ionic core, i.e. perpendicular to the Rydberg orbital plane, which in leading order couples to the $q=0$ component of the quadrupole tensor $\hat{\Theta}_0$ \cite{SM}\nocite{Yoshida2023, Fields2018}. The corresponding interaction Hamiltonian then reads $\hat{H}_Q = -\frac{1}{2} \partial_z E_z \hat{\Theta}_0$, and is diagonal in the fine structure states basis in sufficiently large electric fields \cite{Chen1994}. As a result, the pairs of magnetic substates with $|m_j| = 1/2$ and $3/2$ are shifted in energy by $\langle j,m_j|\hat{H}_Q|j,m_j\rangle = \pm h \delta_n /2$, with $\delta_n = \frac{1}{h} \left| \partial_z E_{z} \right| \Theta_{\text{4\,D}_{3/2}}$ (Fig.~\ref{Fig1}(b)) \cite{SM}. Here, $\Theta_{\text{4D}_{3/2}}$ denotes the quadrupole moment of the 4\,D$_{3/2}$ level and $h$ is Planck's constant. The interaction shift is positive (negative) for $|m_j|=3/2$ ($|m_j|=1/2$), which arises from the oblate (prolate) shape of the orbital and consequently a positive (negative) sign of the quadrupole moment. The strong dependence on the principal quantum number $\delta_n \propto n^{-6}$ comes from the rapid drop of the field gradient $\partial_z E_{z}\propto r_n^{-3}$ with the mean radius $r_n = a_0 n^2$ of the CRS ($a_0$ represents the Bohr radius).
	
	In our experiments, we prepare very high-$n$ CRS $\ket{79\text{C}}$ ($n=79$, $l=m=78$) starting from a 1D chain of ten optical tweezers loaded stochastically with single $^{88}$Sr atoms. Tweezer loading and in-trap laser cooling is described in Ref.~\cite{Holzl2023}. One of the two valence electrons of each is promoted to the circular Rydberg orbit via an initial three-photon optical excitation to the $79^{1}$F$_{3}$ level with lasers at about $\unit[461]{nm}$, $\unit[768]{nm}$, and $\unit[893]{nm}$. The Rydberg excitation is performed in the presence of a magnetic field $B = \unit[0.40(5)]{G}$ pointing along the $z$-direction. This is followed by a radio-frequency (RF) driven adiabatic rapid passage (ARP) to $|79\text{C}\rangle$, using a $\sigma^{+}$-polarized RF field applied on four ring-shaped electrodes surrounding the tweezer array (Fig.~\ref{Fig1}(c)). The circular Rydberg electron is orbiting in the $x$-$y$-plane and is stabilized by an electric field of approximately $\unit[2]{V/cm}$ applied along $z$ using another pair of electrodes made from glass plates coated with indium tin oxide (ITO) which are mounted below and above the ring structure. The plates form a capacitor, which suppresses black-body radiation below a cut-off frequency set by the plate distance ($d \approx \unit[10]{mm}$). This enhances the lifetime by a factor of 8.4  compared to the free-space decay at room temperature, leading to $\unit[2.55(10)]{ms}$ lifetime for $\ket{79\text{C}}$, and allows for high-contrast coherent probing over hundreds of microseconds required to measure the fine quadrupole shift \cite{Hoelzl2024}. We implement a long-lived circular Rydberg qubit by coupling $|79 \text{C}\rangle$ to $|77\text{C}\rangle$ using a two-photon transition at about $\unit[13.9]{GHz}$ as shown in Fig.~\ref{Fig1}(d). The qubit states are then detected via state-selective field-ionization and ion detection on a microchannel plate  (MCP) detector. An exemplary histogram of the ion counts over their arrival time is shown in the inset of Fig.~\ref{Fig1}(c), and illustrates the detection windows for $|79\text{C}\rangle$ and $|77\text{C}\rangle$, in which we count ion events that yield the state populations $p_{79}$ and $p_{77}$, respectively. For details on the experimental sequence for circular state preparation and readout, see Ref.~\cite{Hoelzl2024}.
	
	Once the electron is in the CRS, we can resonantly drive the 5\,S$_{1/2}$ to 5\,P$_{1/2}$ transition of the Sr$^+$  ionic core, using a laser beam at about $\unit[422]{nm}$ (\textit{cf.} Fig.~\ref{Fig1}(b)), linearly polarized along the $y$-direction, i.e. perpendicular to the quantization $z$-axis. Importantly, the resonant excitation of the core electron does not lead to autoionization of the circular state, because of the vanishing spatial overlap of the $|79\text{C}\rangle$ with the 5\,P$_{1/2}$ orbital. Quite differently, if we do not perform the ARP and thus leave the Rydberg electron in the $79^{1}$F$_{3}$ state, rapid autoionization occurs with a 5\,P$_{1/2},79^{1}$F$_{3}$ lifetime of only $\unit[108(19)]{ps}$ \cite{SM}. The absence of CRS autoionization allows us to shelve the ionic core electron into the metastable 4\,D$_{3/2}$ level. The branching ratio for the decay from 5\,P$_{1/2}$ to 4\,D$_{3/2}$ is about 1:17 \cite{NIST_ASD}, and for the polarization used here, we expect equal population of all magnetic substates. Additionally, a second co-propagating laser beam at about $\unit[1092]{nm}$, also linearly polarized along $y$, allows for repumping of all states back into 5\,S$_{1/2}$.
	
	\begin{figure}[!t]
		\centering
		\includegraphics[width=\columnwidth]{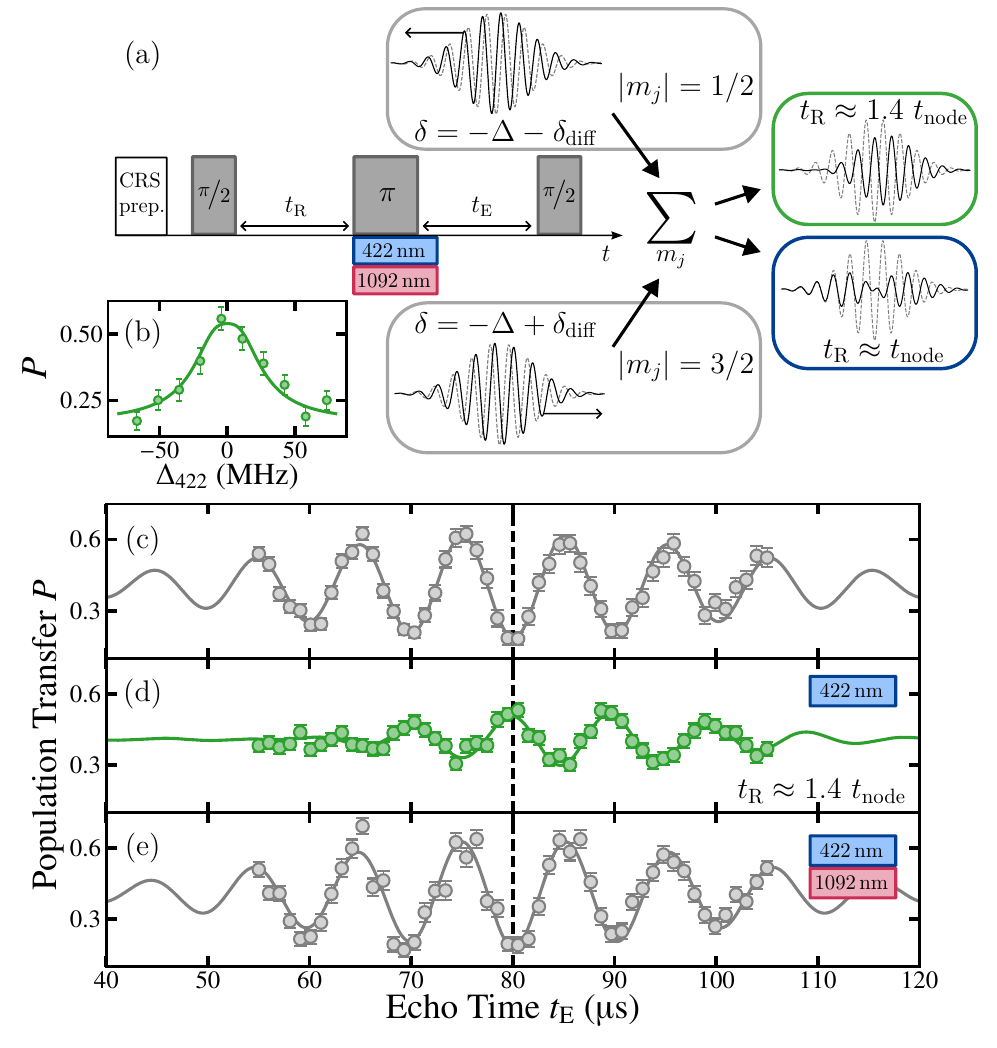}
		\caption{(a) Schematic of the quadrupole shift measurement, including state preparation via ARP, followed by microwave Ramsey interferometry on the circular Rydberg qubit in a Hahn-echo configuration. Laser pulses at $\unit[422]{nm}$ (and $\unit[1092]{nm}$), in parallel with the echo $\pi$-pulse, control the 4\,D$_{3/2}$ population, and thus switch on (and off) the quadrupole interaction in the second interferometer arm, causing a $|m_j|$-dependent shift of the echo signal. Note, that the Gaussian envelope of the signal remains unchanged. The experiment averages over the two possible $|m_j|$-values, generating a beat in the echo signal with a first node at $t_{\text{node}} = 1/(4\delta_{\text{diff}})$. Insets (right) show the resulting signal for $t_\text{R} \approx 1.4 ~ t_\text{node}$ (green) and $t_\text{R} \approx t_\text{node}$ (blue). The latter directly yields the differential quadrupole shift (Fig.~\ref{Fig3}). (c)-(e) Echo signal showing the population transfer $P= p_{77}/(p_{79} + p_{77})$ as a function of $t_{\text{E}}$ for $t_\text{R} =\unit[80]{\muup s}$ without ionic core excitation (c), with the $\unit[422]{nm}$ pulse causing full population transfer to 4\,D$_{3/2}$ (d), and with an additional repumper at $\unit[1092]{nm}$ to 5\,P$_{1/2}$  applied (e). The phase at the maximum echo contrast ($t_{\text{E}}=t_{\text{R}}$, indicated by the dashed line) is shifted by $\pi$ in the presence of the quadrupole interaction. Solid lines are fits of Eq.~\eqref{eq:beat} to the data. (b) Measurement of the 5\,S$_{1/2}$ to 5\,P$_{1/2}$ resonance showing the inversion of the population transfer at $t_{\text{E}}=t_{\text{R}}$ as a function of the $\unit[422]{nm}$ laser detuning $\Delta_{\text{422}}$ for a sequence as in (d). The solid line shows a fit of an optical pumping simulation to extract $\Omega_{422}$ \cite{SM}. Error bars show one standard deviation.
		}
		\label{Fig2}
	\end{figure}
	
	The controlled shelving into 4\,D$_{3/2}$ is used to switch the quadrupole interaction shift on and off, which we calculate to $\delta_{79} = \unit[54.92(32)]{kHz}$ and $\delta_{77} = \unit[64.06(38)]{kHz}$ using the theoretical value $\Theta_{\text{4D}_{3/2}} = \unit[2.029(12)]{e a_0^2}$ \cite{Safronova2008}, and the expectation value of the electric field gradient for $|79\text{C}\rangle$ and $|77\text{C}\rangle$, respectively \cite{SM}. Consequently, the strong $n$-dependence of $\delta_n$ leads to a blue (red) shift, $\delta_{\text{diff}} \equiv (\delta_{77} - \delta_{79})/2$, of our qubit resonance $|79\text{C},4\,\rm{D}_{3/2}\rangle \leftrightarrow |77\text{C},4\,\rm{D}_{3/2}\rangle$ for $|m_j|=1/2$ ($|m_j|=3/2$) with respect to when the ionic core is in the electronic ground state (\textit{c.f.} Fig~\ref{Fig1}(d)). For the high principal quantum numbers studied here, this shift is less than $\unit[5]{kHz}$. Thus, we require more than an order of magnitude longer interrogation times compared to previous measurements in an atomic beam at lower values of $n\approx50$ \cite{Muni2022}, where the circular qubit resonance is shifted by about $\unit[100]{kHz}$. In addition to using laser-cooled and trapped atoms, we achieve this by measuring the quadrupole shift in a circular state Ramsey interferometer complemented by spin-echo to counteract reversible sources of qubit dephasing that appear on shorter timescales.
	
	Our measuring principle is illustrated in Fig.~\ref{Fig2}(a). Following the preparation of $|79\text{C}\rangle$, the Ramsey sequence starts with a typically $\unit[200]{ns}$ long microwave $\nicefrac{\pi}{2}$-pulse to drive the circular qubit. After a free evolution time $t_{\text{R}}$, we apply a $\pi$-pulse to induce a Hahn echo. The interferometer is closed after another wait time $t_{\text{E}}$ with a second $\nicefrac{\pi}{2}$-pulse, and subsequent detection of the population transfer $P= p_{77}/(p_{79} + p_{77})$. For all measurements, we set the microwave to a fixed red detuning of ${\Delta_{\text{}} \approx  \unit[100]{kHz}}$ with respect to the qubit resonance $|79\text{C},5\,S_{1/2}\rangle \leftrightarrow |77\text{C},5\,S_{1/2}\rangle$, defining the fringe period of the echo measurement. The reversible qubit coherence time measured in this work is $T_2^* = \unit[23(2)]{\muup s}$, limited by the first-order magnetic and second-order electric sensitivity of the transition \cite{Hoelzl2024}, which necessitates the echo protocol to measure the small energy shifts from the ionic core \footnote{Note, that this value was increased to $T_2^* =\unit[43(2)]{\muup s}$ in Ref.~ \cite{Hoelzl2024} after careful optimization of the electric and magnetic fields.}. We switch on the quadrupole interaction during $t_{\text{E}}$ by applying a short pulse of the $\unit[422]{nm}$ laser ($\unit[8]{\muup s}$ pulse length and Rabi frequency $\Omega_{422} = \unit[2\pi \times 6.5(10)]{MHz}$ \cite{SM}), starting simultaneously with the echo $\pi$-pulse. The emerging shift on the circular qubit transition then leads to a faster (slower) phase evolution of the qubit for $|m_j|=1/2$ ($|m_j|=3/2$). Accordingly, the echo signal acquires an interaction-induced phase $\phi= \pm 2 \pi \delta_{\text{diff}} t_{\text{E}}$, which shifts the fringes of the echo signal to earlier (later) times for $|m_j|=1/2$ ($|m_j|=3/2$) while the Gaussian envelope remains unchanged (\textit{cf.} Fig.~\ref{Fig2}(a)).
	
	\begin{figure}[!t]
		\centering
		\includegraphics[width=\columnwidth]{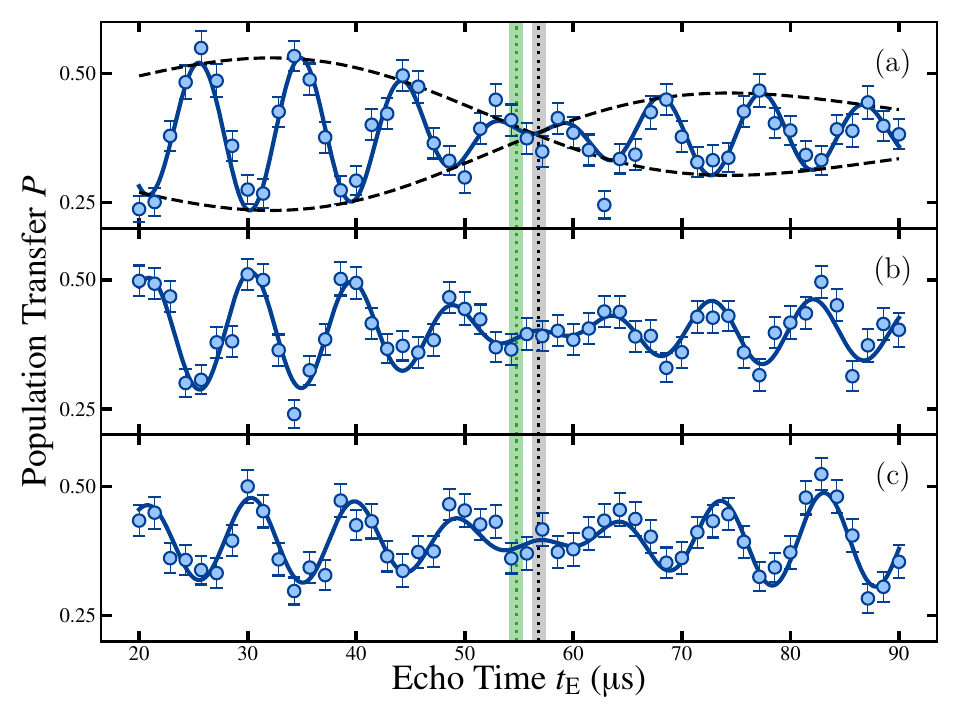}
		\caption{Measurement of the quadrupole interaction shift from beat nodes in the Hahn echo signal. (a)-(c) Echo signals showing the population transfer $P$ around the beat node expected at $t_{\text{node}}= 1/(4\delta_{\text{diff}})$ for $t_{\text{R}}=\unit[50]{\muup s}$ (a), $\unit[55]{\muup s}$ (b), and $\unit[65]{\muup s}$ (c) (\textit{cf}. blue inset Fig.~\ref{Fig2}(a)). Solid lines are fits to the data based on Eq.~\eqref{eq:beat} to extract the quadrupole shift. Vertical lines depict the experimental (black) and theoretical (green) value for $t_{\rm{node}}$, with shaded regions indicating 1$\sigma$ confidence intervals. The difference is attributed to the approximately $\unit[1.7]{\muup s}$ optical pumping time into 4\,D$_{3/2}$. The dashed curve in (a) shows the product of the Gaussian and the beat envelope, $C_{\text{e}}(t_{\rm{E}}) \cdot \text{cos} \left( 2 \pi \delta_{\text{diff}} t_{\rm{E}} \right)$. Error bars denote one standard deviation.
		}
		\label{Fig3}
	\end{figure}
	
	In Fig.~\ref{Fig2}(c), we show the measured echo signal without shelving into 4\,D$_{3/2}$, i.e. the second valence electron remains in 5\,S$_{1/2}$ throughout the measurement sequence. As expected, maximum rephasing, or minimal population transfer $P$, is observed when the two arms of the Ramsey interferometer have the same length $t_{\text{R}}=t_{\text{E}}=\unit[80]{\upmu s}$. In the case of adding the $\unit[422]{nm}$ pulse to switch on the quadrupole interaction (Fig.~\ref{Fig2}(d)), we find a complete $\pi$-phase shift together with a reduced contrast. Despite that, the expected phase acquired when considering a single $m_j$ level is only $\approx 0.73\,\pi$. Indeed, our shelving pulse does not populate a single $m_j$ level but rather all 4\,D$_{3/2}$ substates with equal probability. Consequently, each realization of the Ramsey sequence on an individual atom leads to one of the two scenarios described above, with an echo that is phase-shifted by the same amount but with opposite signs for $|m_j|=1/2$ and $|m_j|=3/2$. One then detects the statistical average of the two outcomes, which for the conditions in Fig.~\ref{Fig2}(d) yields the observed $\pi$-shifted sinusoidal fringe pattern. Additionally, measuring the Ramsey signal at $t_{\text{E}}=t_{\text{R}}$ as a function of the detuning of the $\unit[422]{nm}$ laser (Fig.~\ref{Fig2}(b)) allow us to identify the 5\,S$_{1/2}$ to 5\,P$_{1/2}$ resonance via the quadrupole shift. Note, that the value for $\Omega_{422}$ given above is obtained from fitting a simulation of the optical pumping dynamics to the spectrum (solid line in Fig.~\ref{Fig2}(b)) \cite{SM}.

	In a next step, we repeat the same protocol but also apply the $\unit[1092]{nm}$ laser on resonance with the 4\,D$_{3/2}$ to 5\,P$_{1/2}$ transition together with the $\unit[422]{nm}$ pulse (Fig.~\ref{Fig2}(e)). The set laser power effectively pumps the 4\,D$_{3/2}$ population back into the 5\,S$_{1/2}$ ground state. Consequently, we again observe the same signal as in Fig.~\ref{Fig2}(c). However, now the ionic core electron scatters about 100 photons during the laser pulse \cite{SM}, without evidence for loss of coherence on the circular Rydberg qubit.
	
	We now turn to an accurate measurement of the quadrupole interaction shift. Incoherent summing over the two possible outcomes of the Ramsey sequence discussed above predicts a beat in the interferometry signal (\textit{cf.} Fig.~\ref{Fig2}(a)). The beat frequency $\delta_{\text{diff}}$ provides a direct and precise measure of the quadrupole-induced differential shift on the qubit resonance. For our parameters, we expect the first node at the time $t_{\text{E}} = t_{\text{node}} = 1/(4\delta_{\text{diff}}) \approx \unit[55]{\muup s}$. This node is already visible in the data of Fig.~\ref{Fig2}(d), and the observed $\pi$-shift of the maximum echo at $t_{\text{E}}=t_{\text{R}}$ is actually a robust feature that appears for times $t_\text{R} >  t_{\text{node}}$. In order to optimize the signal contrast around the expected node position, we repeat the measurement of Fig.~\ref{Fig2}(d), but now set $t_{\text{R}}$ to three different values around $t_{\text{node}}$. The results are shown in Fig.~\ref{Fig3}, and reveal that the node position is indeed robust and independent of $t_{\text{R}}$. The signal follows the functional form
	
	\begin{align}
		P = C_{\text{e}}(t_{\rm{E}}) ~  \text{cos}\left(2 \pi \Delta (t_{\rm{E}} - t_{\rm{R}}) \right) \text{cos} \left( 2 \pi \delta_{\text{diff}} t_{\rm{E}} \right)  + c_\text{off}, \label{eq:beat}
	\end{align}
	with an offset $c_\text{off}$, the microwave detuning $\Delta$ and a Gaussian envelope $C_{\text{e}}=\text{exp}(-(t_{\rm{E}}-t_{\text{R}})^2/2 T_2^*{^2})$ accounting for the reversible qubit dephasing time $T_2^*$. Fitting Eq.~\eqref{eq:beat} to the data shown in Fig.~\ref{Fig3} yields a weighted mean of $\delta_{\text{diff}} = \unit[4.40(5)]{kHz}$. From this we can then deduce the quadrupole moment to $\Theta_{\text{4D}_{3/2}} = \unit[1.98(\pm 0.02)_\text{stat.} (^{+0.06}_{-0.00})_\text{syst.}]{e a_0^2}$, which is in good agreement with the calculated value in Ref.~\cite{Safronova2008}. In addition to statistical uncertainty, the uncertainty includes a systematic error resulting from the finite time required to optically pump into the 4\,D$_{3/2}$ state, which consequently shifts $t_\text{node}$ to larger times (\textit{cf.} vertical lines in Fig.~\ref{Fig3}). This time is estimated to be $t_{\text{pump}} \approx \unit[1.7]{\upmu s}$ \cite{SM}.

	\begin{figure}[!t]
		\centering
		\includegraphics[width=\columnwidth]{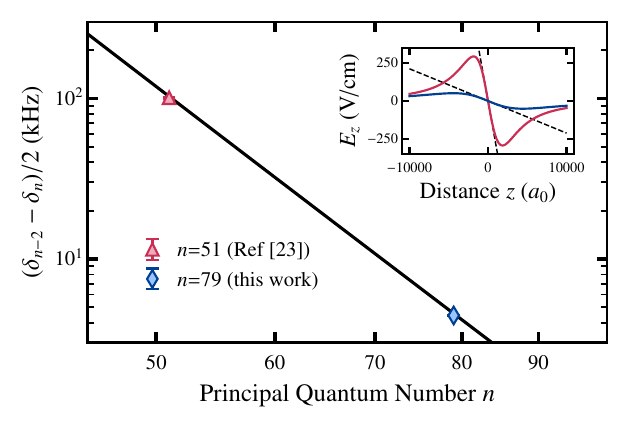}
		\caption{Quadrupole shift on the circular Rydberg qubit, $(\delta_{n-2}-\delta_{n})/2$, as a function of principal quantum number $n$, comparing data from this work at $n=79$ with an atomic beam experiment at $n=51$ \cite{Muni2022}. The solid line shows the calculated differential shift \cite{SM}. Note, the axes are both scaled logarithmically. The inset depicts the electric field generated by the Rydberg electron along the $z$-axis for the two values of $n$. Dashed lines indicate the large difference in the electric gradient $\partial_z E_z$ at $z = 0$, which scales with $n^{-6}$ and is responsible for the quadrupole interaction. Error bars are smaller than the data points.}
		\label{Fig4}
	\end{figure}
	
	Finally, we confirm the theoretically predicted $n^{-7}$ scaling of $\delta_{\text{diff}}$ in Fig.~\ref{Fig4} over more than an order of magnitude by comparing our results at very high-$n$ with previous measurements in an atomic beam at $n=51$ \cite{Muni2022}. The comparison directly illustrates the large reduction in accessible energy scales for electron-electron interactions in doubly excited Rydberg systems enabled by our work.
	
	In conclusion, we implemented quadrupole coupling between very high-$n$ circular Rydberg qubits and metastable ionic core states prepared from an optical tweezer array. Our work shows how ionic-core-induced interaction shifts of only a few kilohertz on Rydberg qubits can be precisely sensed by exploiting up to millisecond-long interrogation times, elusive to low-$l$ Rydberg states. Combining laser cooling and atom trapping, circular state lifetime enhancement in a black-body radiation suppression capacitor, and spin-echo qubit interrogation, was key to access the fine energy shifts in a room temperature setup. Our results open up exciting routes to use the optically active core of circular Rydberg atoms for quantum applications. The observation that a hundred scattered photons on the ionic core do not deteriorate the coherence of the circular qubit paves the way for laser cooling and non-destructive fluorescence imaging of Rydberg atoms in the context of neutral atom quantum simulation. The quadrupole shift can further be utilized to implement local optical control of circular state qubits, thereby bringing together coherent microwave control with the rich optical toolbox for trapped ion qubits.

	\begin{acknowledgments}
		We thank the \href{https://thequantumlaend.de/}{\textsc{Quantum Länd}} team and Tilman Pfau for fruitful discussions and Jennifer Krauter for proofreading. We acknowledge funding from the Federal Ministry of Education and Research (BMBF) under the grants CiRQus and QRydDemo, the Carl Zeiss Foundation via IQST, the Horizon Europe programme HORIZON-CL4-2021-DIGITAL-EMERGING-01-30 via the project 101070144 (EuRyQa), and the Vector Foundation.
	\end{acknowledgments}

\clearpage

	\section{Supplementary Material: Coherent coupling of circular Rydberg qubits to inner shell excitations}
	
	\subsection{Rydberg F-state Autoionization}
	\label{sec_app_ai}
	
\begin{figure}[!ht]
		\centering
		\includegraphics[width=\columnwidth]{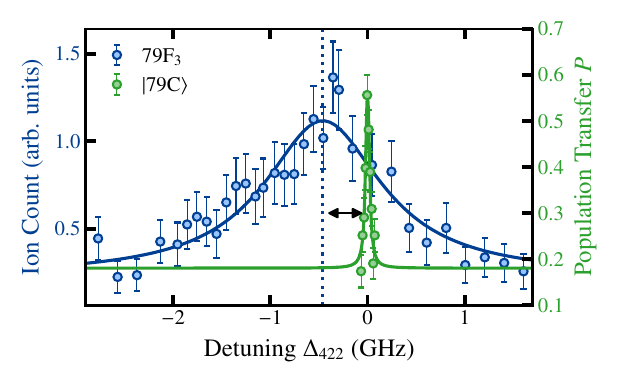}
		\caption{Autoionization resonance of the 5\,P$_{1/2},79^{1}$F$_{3}$ state (blue circles) as a function of the $\unit[422]{nm}$ laser detuning. The data from Fig.~2(b) of the main article (green circles), showing the 5\,S$_{1/2}$ to 5\,P$_{1/2}$ resonance with the Rydberg electron in the $\ket{79{\rm{C}}}$ circular state, is plotted for comparison. Zero detuning is set to the resonance position of the $\ket{79{\rm{C}}}$ data. The dashed line indicates the fitted center of the 5\,P$_{1/2},79^{1}$F$_{3}$ spectrum, which is shifted from $\Delta_{422}=0$ as marked by the black arrow. The blue solid line is a Lorentzian fit to the 5\,P$_{1/2},79^{1}$F$_{3}$ data. The green solid line depicts the fit of the optical pumping model from Fig.~2(b) for the $\ket{79\rm{C}}$ data. The data ordinate is given by the correspondingly colored axis.
		}
		\label{Fig5}
	\end{figure}

	As stated in the main article, we observe rapid autoionization of the $79^{1}$F$_{3}$ Rydberg state when driving the 5\,S$_{1/2}$ to 5\,P$_{1/2}$ transition of the Sr$^+$  ionic core. Figure \ref{Fig5} shows the measured autoionization resonance. For this measurement, we pulse the \unit[422]{nm} laser directly after exciting the $79^{1}$F$_{3}$ state without applying the ARP to the circular state. The created free ion is then guided to the MCP with an electric field ($\approx \unit[7.8]{V/cm}$) much too weak to ionize the Rydberg level. A scan of the \unit[422]{nm} laser frequency reveals a Lorentzian line shape with a full width at half maximum of $\rm{FWHM}_{\rm{AI}} = \unit[1.47(22)]{GHz}$, from which we obtain the lifetime of the autoionizing state $\tau_{\rm{AI}} = 1/(2 \pi \times \rm{FWHM}_{\rm{AI}}) =  \unit[108(19)]{ps}$. This value seems rather high compared with previous measured values \cite{Yoshida2023_S, Fields2018_S}.
	We attribute this discrepancy to the fact that the isolated core excitation is performed in the extraction electric field, which mixes the F-state with states with higher $l$, decreasing the autoionization rate. Although we have taken care to keep the $\unit[422]{nm}$ laser power low enough to avoid saturation and line broadening, we cannot fully exclude residual systematics on the measured $\tau_{\rm{AI}}$ to smaller values.
	
	When comparing this to the 5\,S$_{1/2}$ to 5\,P$_{1/2}$ resonance from the data of Fig.~2(b) of the main article, for which the Rydberg electron is in the circular state, one also finds a red shift of $\unit[460(51)]{MHz}$ of the $79^{1}$F$_{3}$ autoionization resonance position. This shift is attributed to the difference in quantum defects for the 5\,S$_{1/2},79^{1}$F$_{3}$ and 5\,P$_{1/2},79^{1}$F$_{3}$ state, and is found in fair agreement with measurements at high values of $n$ reported in Ref.~\cite{Yoshida2023_S}.
	
	\subsection{Optical Pumping Simulation}
	\label{sec_app_opticalpumping}
	
	We model the optical pumping from 5\,S$_{1/2}$ into the metastable 4\,D$_{3/2}$ level via the 5\,P$_{1/2}$ state by a three-level system. To this end, we calculate the time evolution of the density matrix by solving a Lindblad equation for the $\unit[8]{\muup s}$ long \unit[422]{nm} laser pulse applied in the experiment and extract the final population in 4\,D$_{3/2}$ as a function of $\Delta_{422}$. The obtained line shape, which due to saturation of the shelving process on resonance is no longer Lorentzian, is fitted to the data in Fig.~2(b) of the main article. Decay rates $\Gamma_{422}=\unit[2 \pi \times 20.36]{MHz}$ ($\Gamma_{1092}=\unit[2 \pi \times 1.19]{MHz}$) from 5\,P$_{1/2}$ into 5\,S$_{1/2}$ (4\,D$_{3/2}$) are taken from \cite{NIST_ASD_S}. From the fit (solid line in Fig.~2(b)), we obtain $\Omega_{422}=\unit[2\pi \times 6.5(10)]{MHz}$, which corresponds to a saturation parameter $s_{422}=2 \Omega_{422}^2 / \Gamma_{422}^2 = 0.21(2)$. For these parameters, the calculated 1/e-time for pumping into 4\,D$_{3/2}$ is $t_{\rm{pump}} \approx \unit[1.7]{\muup s}$. Accordingly, the line in Fig.~2(b) is broadened with respect to the natural linewidth for the comparatively long pulse time of the \unit[422]{nm} laser.
	
	From the same model, we also extract the estimated number of scattered \unit[422]{nm} photons for the scenario of Fig.~2(e) in the main article, where the \unit[1092]{nm} repumper is switched on simultaneously with the \unit[422]{nm} pulse. For the \unit[1092]{nm} power used in the experiment, we effectively saturate the repumper transition ($s_{1092} \gtrsim 1400$). Consequently, the resonant \unit[422]{nm} scattering rate is $\Gamma_{422} s_{422} / (2(s_{422}+1)) = \unit[2 \pi \times 1.8(2)]{MHz}$, which results in about 100 scattered photons for the $\unit[8]{\muup s}$ laser pulse length.
	
	\subsection{Calculation of the Quadrupole Shift}
	\label{sec_app_calc}
	
	For calculating the quadrupole interaction shift on the $|n{\rm{C}}, 4\,{\rm{D}}_{3/2} \rangle$ $m_j$-levels, we follow the multipole expansion described in Refs.~\cite{Muni2022_S, muni_thesis_S}, and specifically evaluate the leading-order term describing the coupling of the quadrupole moment of $4\,{\rm{D}}_{3/2},m_j$ to the electric field gradient produced by the Rydberg electron at the ionic core. The Hamiltonian describing this coupling, $\hat{H}_Q = \nabla {\hat{\mathbf{E}}}^{(2)} \cdot \hat{\Theta}^{(2)}$, contains the rank-2 tensor operators for the electric field gradient $\nabla {\hat{\mathbf{E}}}^{(2)}$ acting on the Rydberg electron and the quadrupole moment $\hat{\Theta}^{(2)}$ acting on the core electron.
	
	The corresponding sum over the tensor components entering the matrix elements
	$$
	\langle n {\rm{C}}, m_j| \hat{H}_Q |n {\rm{C}}, m_j'\rangle  = \sum_{q=-2}^{+2} \langle n {\rm{C}}| \nabla \hat{E}^{(2)}_q |n {\rm{C}} \rangle \langle m_j| \hat{\Theta}^{(2)}_q |m_j', \rangle
	$$
	reduces to a single diagonal term for the $q=0$ component
	$$
	\langle n {\rm{C}}, m_j| \hat{H}_Q |n {\rm{C}}, m_j\rangle  = \langle n {\rm{C}}| \nabla \hat{E}^{(2)}_0 |n {\rm{C}} \rangle \langle m_j| \hat{\Theta}^{(2)}_0 |m_j \rangle
	$$
	because $\langle n {\rm{C}}| \nabla \hat{E}^{(2)}_q |n {\rm{C}}\rangle =0 $ for $|q|>0$. This step assumes truncation of the basis for the Rydberg electron to the circular state only, and thus neglects admixing of elliptical states by the electrostatic interaction between the two electrons \cite{Chen1994_S}. For our experiments, this is justified due to the large energetic separation of those states caused by the applied magnetic and electric field. Specifically, the spacing to the state closest in energy to $79 {\rm{C}}$ is $\approx \unit[6.5]{MHz}$, about two orders of magnitude larger than $\delta_{79}$.
	
	The fact that the interaction shift is equal in magnitude but opposite in sign for $|m_j|=1/2$ and $3/2$ originates from the Clebsch-Gordon coefficient when reducing the matrix element $\langle m_j| \hat{\Theta}^{(2)}_0 |m_j \rangle = \langle j m_j 2 0 | j m_j\rangle \langle j \| \hat{\Theta}^{(2)} \| j\rangle$, giving
	$$
	\langle j m_j 2 0 | j m_j\rangle =  \left\{
	\begin{array}{rcl}
		-\sqrt{\frac{1}{5}} & \text{for} & |m_j|=1/2 \\
		+\sqrt{\frac{1}{5}} & \text{for} & |m_j|=3/2
	\end{array}
	\right.
	$$
	for the angular momentum $j=3/2$ of the $4\,{\rm{D}}_{3/2}$ state. The theoretical value of the quadrupole moment $\Theta_{\text{4D}_{3/2}} = \unit[2.029(12)]{e a_0^2}$ reported in the main text is defined in Ref.~\cite{Safronova2008_S} as the $q=0$ component of the quadrupole operator for the magnetic substate with maximum projection $m_j=j$, and thus
	$$
	\langle m_j| \hat{\Theta}^{(2)}_0 |m_j \rangle = \left\{\begin{array}{rcl}-\Theta_{\mathrm{4D}_{3/2}} &\rm{for}& |m_j|=1/2\\+\Theta_{\mathrm{4D}_{3/2}} &\rm{for}& |m_j|=3/2\end{array}\right. \, .
	$$
	
	Finally, obtaining values for
	\begin{equation}
		\delta_n = \frac{1}{h} \left| \frac{\partial E_{z}}{\partial z} \right| \Theta_{\mathrm{4D}_{3/2}},
		\label{eq:apx_delta}
	\end{equation}
	as defined in the main text, requires calculating the $n$-dependent expectation value $\langle n {\rm{C}} |\partial_z E_{z}| n {\rm{C}} \rangle$. Using $\partial_z E_{z} = -2 \nabla \hat{E}_0^{(2)}$ and expressing $\nabla \hat{E}_0^{(2)}$ in terms of spherical harmonics, allows for separating the integral
	\begin{equation}
		|\partial_z E_{z}|= 2 \langle n {\rm{C}} | \nabla \hat{E}_0^{(2)}| n {\rm{C}} \rangle = \frac{2 E_\mathrm{h}}{e a_0^2} \cdot I_r \cdot I_a
	\end{equation}
	into a radial ($I_r$) and an angular ($I_a$) part, with $E_\mathrm{h}$ being the Hartree energy.
	By solving the separated integrals, one obtains
	\begin{align}
		I_r &= \frac{1}{n^3 ( l(l+1/2)(l+1))} \quad \text{and} \notag \\
		I_a &= - (-1)^{m} \cdot (2l+1) \cdot
		\begin{pmatrix}
			l & 2 & l \\
			0 & 0 & 0
		\end{pmatrix}
		\begin{pmatrix}
			l & 2 & l \\
			-m & 0 & m
		\end{pmatrix} \, . \notag
	\end{align}
	
	Here, $\left(\cdot\right)$ denotes the 3j-symbol.
	Finally, with $m=l=n-1$, the quantum numbers of the circular Rydberg electron orbital, Eq.~\eqref{eq:apx_delta} results in a compact form
	\begin{equation}
		\delta_n = \frac{E_\mathrm{h}}{e a_0^2 h} \frac{4}{4n^6-n^4} \Theta_{\mathrm{4D}_{3/2}},
	\end{equation}
	from which one obtains the values stated in the main article, $\delta_{79} =\unit[54.92(32)]{kHz}$ and $\delta_{77} =\unit[64.06(38)]{kHz}$. Here, the uncertainty stems from the uncertainty of the theoretical value for $\Theta_{\mathrm{4D}_{3/2}}$ given in Ref.~\cite{Safronova2008_S}.

	\clearpage
	
\end{document}